\documentclass{elsart1p}
\usepackage{graphicx}
\usepackage{amssymb}
\begin{document}
\begin{frontmatter}
%
%
\title{Time Evolution of Thermal Photon Elliptic Flow}
\author{ Rupa Chatterjee, Dinesh K. Srivastava}
\address{Variable Energy Cyclotron Centre, 1/AF, Bidhan Nagar, Kolkata 
700 064, India}\author{Ulrich Heinz}
\address{Physics Department, The Ohio State University, Columbus, 
OH 43210, USA}
\begin{abstract}
Elliptic flow of thermal photons has a great potential to explore 
the early time dynamics and evolution of Quark Gluon Plasma. $p_T$ 
dependent temporal contours for photon spectra and elliptic flow 
from quark matter and hadronic phases, as well as the $p_T$ 
integrated results show gradual build-up of flow with time 
and relative contributions from different phases to that 
very clearly. Unlike hadrons, photon flow is quite sensitive 
to the initial thermalization time $\tau_0$, and its value 
can be estimated with the experimental determination of $v_2$.
\end{abstract}
\begin{keyword}
Photons, elliptic flow, time evolution, $\tau_0$ etc.
\PACS
\end{keyword} 
\end{frontmatter}
\section{Introduction}
\label{first}
The observation of large anisotropic flow or in particular 
elliptic flow of different hadronic species at the Relativistic 
Heavy Ion Collider (RHIC) at Brookhaven National Lab, New York, 
provides a strong confirmation of the formation of QGP. Photons are 
known to probe the properties of QGP in a relatively clean manner 
compared to hadrons. Unlike hadrons, photons are emitted from every 
stage of the expanding system and also carry undistorted information 
from the production point to the detector. Elliptic flow is known 
to  provide information about early thermalization and collectivity 
in the hot and dense matter. The flow parameter $v_2$ is quantified 
as the 2nd Fourier co-efficient of the particle distribution in the 
transverse momentum plane as (at mid-rapidity and for collisions of 
same type of nuclei only even cosine terms survive in the 
series below):
\begin{eqnarray}
\frac{dN(b)}{p_T dp_T \, dy \,d\phi} = \frac{dN(b)}{ 2 \pi p_T \, dp_T 
\, dy}[1 +  2 v_2(p_T,b) \cos(2 \phi)+ 2 v_4(p_T,b) \cos(4 \phi)+ \dots ]
\end{eqnarray}
We have shown that elliptic flow of thermal photons for $200 A$ 
GeV Au+Au collisions at RHIC (using ideal hydrodynamic model)
reflects the anisotropies of the early partonic phase at large 
values of transverse momentum $p_T$~\cite{phot_flow}. In a
recent calculation of photon $v_2$ (considering prompt and jet 
photons along with thermal), it is shown that the thermal $v_2$ 
dominates over others for $p_T < 5$ GeV~\cite{TGFH}. 
%
\begin{figure}
\centering
\includegraphics[height=3.7cm]{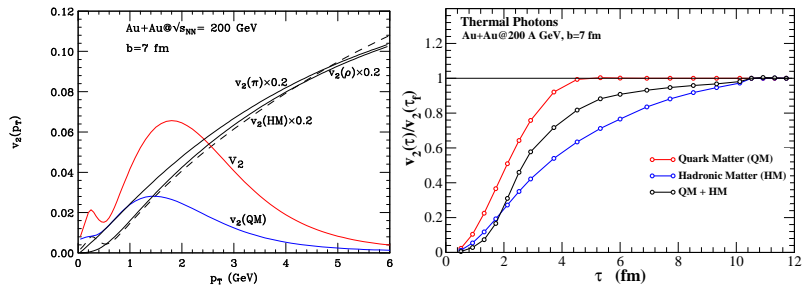}
\includegraphics[height=3.7cm]{v_2_t.eps}
\caption{Left panel: $v_2$ for thermal photons along with 
contributions from QM and HM. Right panel: $p_T$ integrated 
$v_2$ results from different phases as a function of $\tau$.}
\label{fig:1}      
\end{figure}
\section {Time evolution of photon $v_2$}
\begin{figure}
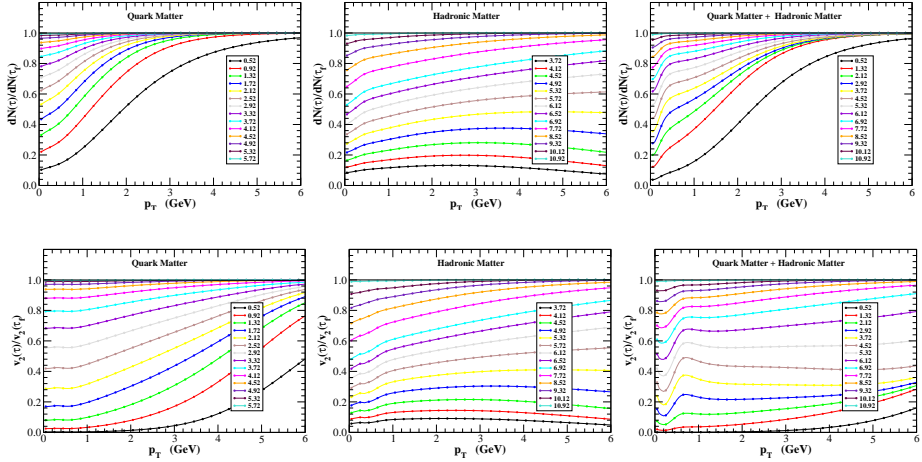

\centering
\includegraphics[height=2.8cm]{qm_dn.eps}
\includegraphics[height=2.8cm]{hm_dn.eps}
\includegraphics[height=2.8cm]{sum_dn.eps}
\vspace{0.45 cm}
\linebreak
\includegraphics[height=2.8cm]{qm_v2.eps}
\includegraphics[height=2.8cm]{hm_v2.eps}
\includegraphics[height=2.8cm]{sum_v2.eps}
\caption{$p_T$ spectra [upper panel] and elliptic flow [lower panel] of 
thermal photons for different values of $\tau$ from QM, HM and the sum 
of the two phases.}
\label{fig:2}     
\end{figure}

The time evolution of photon spectra and $v_2$ is studied for impact 
parameter $b=7$ fm and $\tau_0=0.2$ fm considering the same initial 
conditions as given in Ref.~\cite{phot_flow}. Rates of photon production 
from quark matter and hot hadronic gas are taken from Arnold 
{\it et al.}~\cite{arnold} and Turbide {\it et al.}~\cite{turbide} 
respectively, which provide good description of the single photon data 
at RHIC as well as at SPS energies. 
The $p_T$ dependent elliptic flow parameter for thermal photons along 
with contributions to $v_2$ from quark matter (QM) and hadronic matter 
(HM) phases are shown separately in left panel of Fig.\ref{fig:1}. In 
the right panel of same figure, the $p_T$ integrated $v_2$ from 
different phases as a function of proper time $\tau$ and normalized 
with $v_2(\tau_f)$ from respective phases are shown, where $\tau_f$
( $\approx 12$ fm) is the time when freeze-out is completed~\cite{evolution}. 
As shown in the figure, the $v_2(QM)$ saturates early within a time 
period of $5$ fm. The $v_2(HM)$ is small at the beginning and saturates 
much later compared to $v_2(QM)$. Contours of $p_T$ dependent spectra 
and $v_2$ at various values of  $\tau$ from QM, HM and the sum of 
the two phases are shown explicitly in Fig.~\ref{fig:2}. At large 
$p_T$ ($ > 3$ GeV), most of the photons from QM are emitted very early 
within a time period of $2$ fm. However, the $v_2(QM)$ is not that 
strong at early times and it gradually builds-up as shown in the 
left panel of Fig.~\ref{fig:2}. Photons from HM and their $v_2$, 
both are very small at the beginning and become significant only 
after 4-5 fm. The sum spectra and $v_2$ show clear QM dominance 
at early times or large $p_T$ and the HM contribution is substantial 
only for lower $p_T$ or later times.
\section{$\tau_0$ from elliptic flow of thermal photons}
\begin{figure}
\centering
\includegraphics[height=5.5cm,angle=-90]{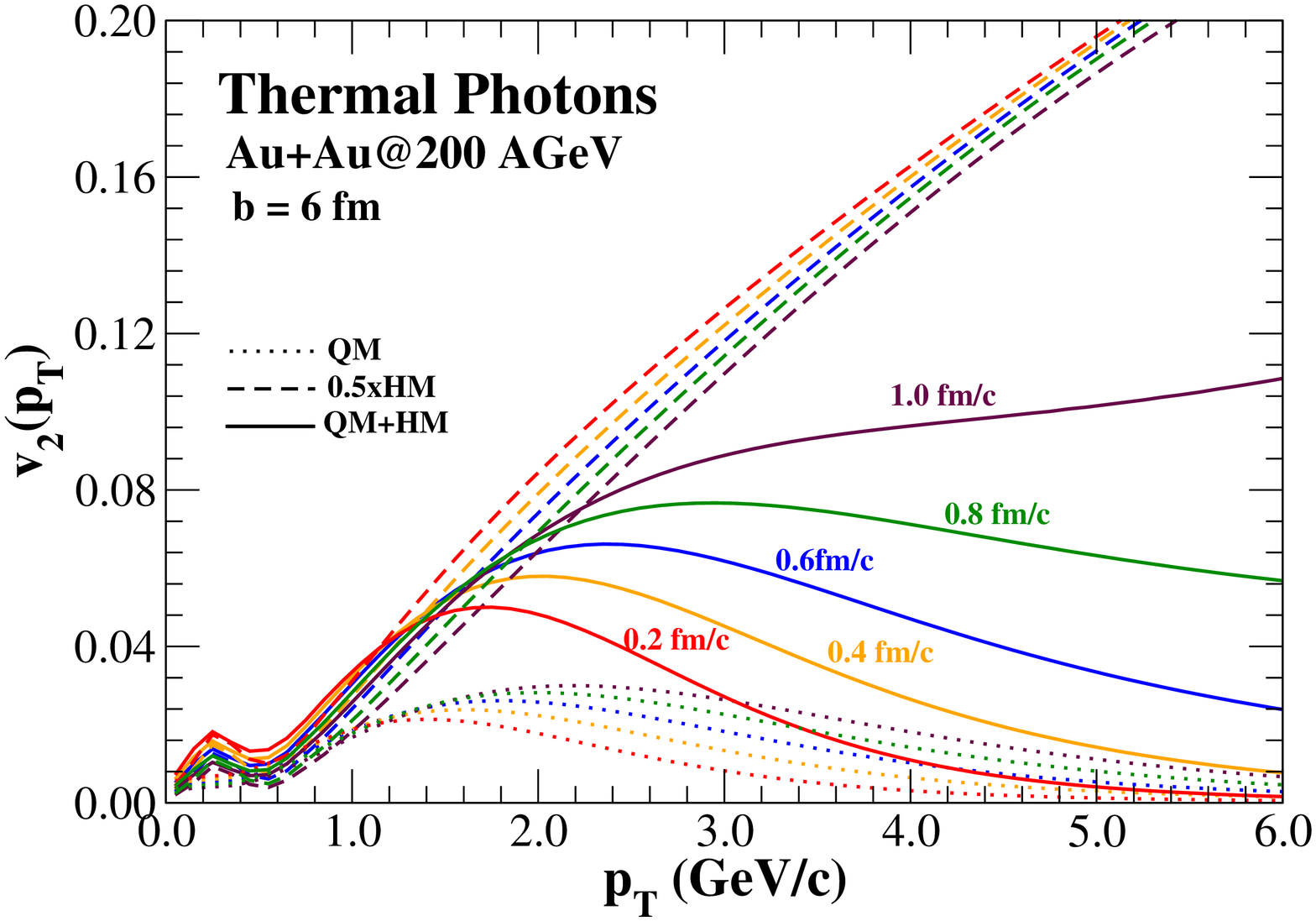}
\includegraphics[height=5.5cm,angle=-90]{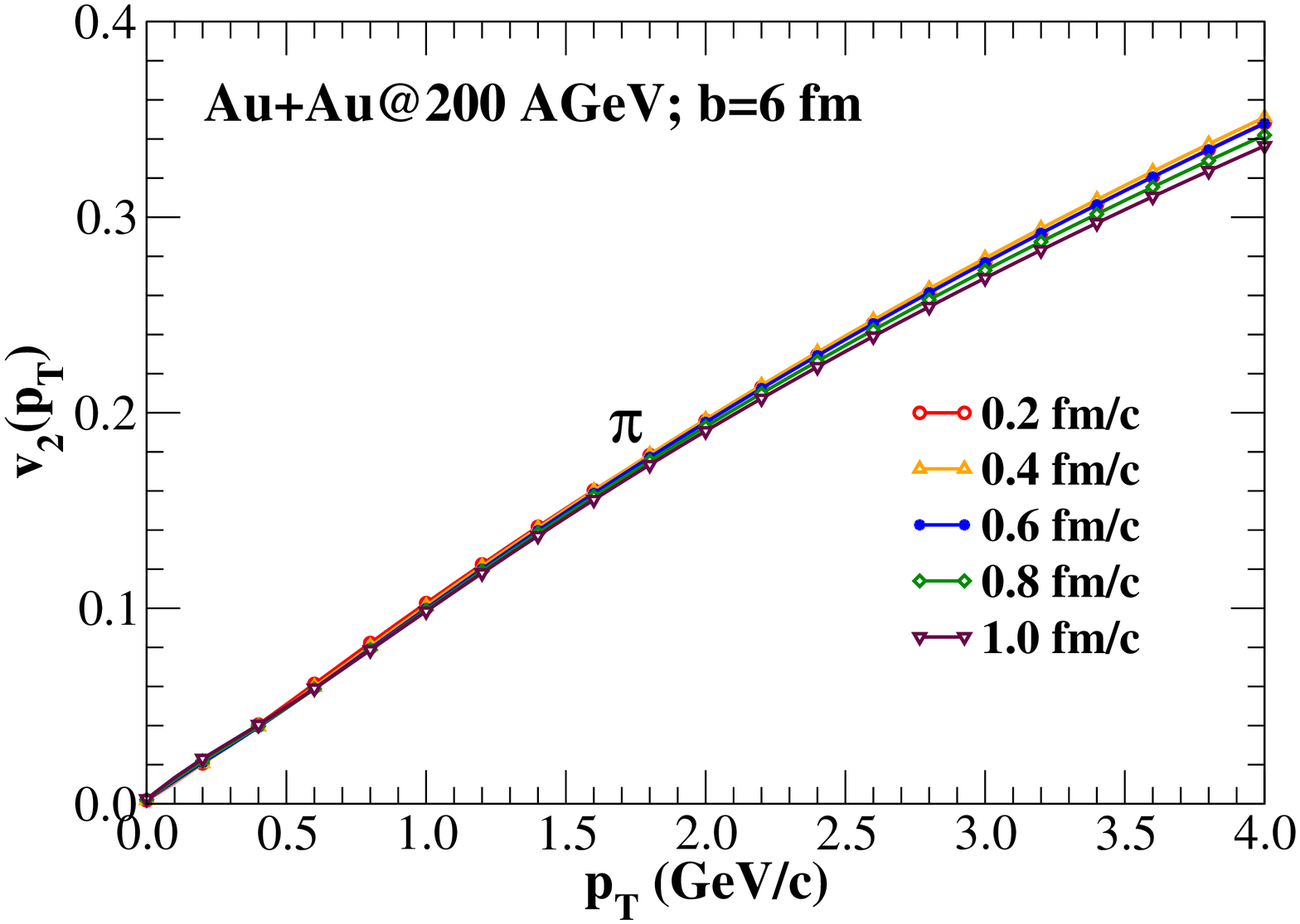}
\caption{ $v_2(p_T)$ for thermal photons [left panel] and $\pi$ mesons 
[right panel] at different $\tau_0$.}
\label{fig:3}  
\end{figure}
The final $v_2(p_T)$ for photons is quite sensitive to 
the initial thermalization time $\tau_0$ for a fixed value of 
particle multiplicity, and the value of $\tau_0$ can thus be estimated 
with the help of experimental results~\cite{tau_rhic}. The differential 
elliptic flow $v_2(p_T)$ for different $\tau_0$ ranging from 0.2 to 1.0 
fm at $b=6$ fm, are shown in left panel of Fig.~\ref{fig:3}. For smaller 
$\tau_0$, the number of photons from QM increases compared to larger 
$\tau_0$ at high $p_T$. As $v_2(QM)$ is smaller at earlier 
times and $v_2(HM)$ does not change significantly with changing 
$\tau_0$, the total $v_2$ decreases with smaller $\tau_0$. As 
hadrons are emitted from freeze-out surface, their $v_2$ is 
affected only marginally with changing values of $\tau_0$ (right panel 
of Fig.~\ref{fig:3}).

In conclusion, the time evolution results for photon spectra and 
elliptic flow show gradual build-up of the flow parameter along with 
relative contributions from different phases with time and at different 
$p_T$. Photon flow is quite sensitive to the value of $\tau_0$, which 
can be estimated by experimental determination of the flow parameter.

\end{document}